\documentclass[epj]{svjour}
\usepackage{graphics}

\def\beq{\begin{equation}}
\def\eeq{\end{equation}}
\def\bea{\begin{eqnarray}}
\def\eea{\end{eqnarray}}
\def\beqa{\begin{equation}\begin{array}{l}}
\def\eeqa{\end{array}\end{equation}}

\def\half{\mbox{\small{$\frac{1}{2}$}}}

\def\barr{\left(\begin{array}{c}}
\def\earr{\end{array}\right)}
\def\bmat{\left(\begin{array}{cc}}
\def\emat{\end{array}\right)}
\def\al{\alpha}
\def\be{\beta}
\def\ga{\gamma} 
\def\de{\delta} \def\De{\Delta}
\def\veps{\varepsilon}  \def\eps{\epsilon}

\def\la{\lambda} \def\La{{\Lambda}}

\def\si{\sigma} \def\Si{{\it\Sigma}}
  \def\Th{\Theta}

\def\pa{\partial}

\def\pa{\partial}

\def\nn{\nonumber}

\def\Tau{{\mathcal T}}
\def\lag{{\mathcal L}}

\def\mathscr{\mathcal}

\def\3d{3-D}

\def\ol#1{\overline{#1}}
\def\amm{a.m.m.}
\def\ceft{$\chi$EFT}

\begin{document}
\title{Two-photon physics}
\author{Marc Vanderhaeghen \inst{1} \inst{2}
}                     
\institute{Physics Department, The College of William \& Mary, 
Williamsburg, VA 23187, USA 
\and 
Theory Group, Jefferson Lab, 12000 Jefferson Ave, 
Newport News, VA 23606, USA}
\date{Received: date / Revised version: date}
%
\abstract{
It is reviewed how Compton scattering sum rules relate 
low-energy nucleon structure quantities to the nucleon 
excitation spectrum. In particular, the GDH sum rule 
and recently proposed extensions of it will be discussed. 
These extensions are sometimes 
more calculationally robust, which may be an advantage when 
estimating the chiral extrapolations of lattice QCD results, 
such as for anomalous magnetic moments. 
Subsequently, new developments in our description of the nucleon 
excitation spectrum will be discussed, in particular 
a recently developed chiral effective field theory 
framework for the $\Delta(1232)$-resonance 
region.  Within this framework, we discuss results on 
$N$ and $\Delta$ masses, 
the $\gamma N \Delta$ transition and the $\Delta$ magnetic dipole moment. 
\\
} 
\maketitle
\section{Introduction}
\label{intro}

Sum rules for Compton scattering off a nucleon offer a unique tool 
to relate low energy nucleon structure quantities to the nucleon 
excitation spectrum~\cite{Drechsel:2002ar}. 
E.g., the Gerasimov, Drell, Hearn (GDH) sum rule (SR)~\cite{GDH} relates a
system's anomalous magnetic moment to a weighted integral over a
combination of doubly polarized photoabsorption cross sections.  
Impressive experimental programs to measure 
these photoabsorption cross-sections for the
nucleon have recently been carried out at ELSA and MAMI (for a
review see Ref.~\cite{Drechsel:2004ki}).  Such measurements
provide an empirical test of the GDH SR, and can be used to
generate phenomenological estimates of electromagnetic
polarizabilities via related SRs.
The GDH SR is particularly interesting because both its left- and
right-hand-sides can be reliably determined, thus providing a
useful verification of the fundamental principles (such as
unitarity and analyticity) which go into its derivation.  At the
present time, it has been established that 
the proton sum rule is satisfied within the
experimental precision, while the case is still out for
the neutron.  

After a lightning review of the GDH and related sum rules
in Sec.~\ref{sec:1}, I discuss a recently proposed 
linearized  version of the 
GDH sum rule~\cite{Pascalutsa:2004ga,Holstein:2005db}.
When applying this new sum rule to the nucleon in the context of chiral
perturbation theory, it allows for an
elementary calculation (to one loop) of quantities such as
magnetic moments and polarizabilities to all
orders in the heavy-baryon expansion.  The chiral behavior of the
nucleon magnetic moment allows to make a link with lattice QCD calculations. 

Subsequently, the nucleon excitation spectrum 
is discussed in Sec.~\ref{sec:2}. 
Many Compton scattering sum rules, such as the GDH sum rule, are 
dominated by the $\Delta$(1232) resonance. I discuss a 
recently proposed relativistic chiral effective field theory 
as a new systematic framework to both extract resonance properties
from the experiment and to perform a chiral extrapolation of lattice 
QCD results for those resonance properties.

\section{Sum rules in Compton scattering}
\label{sec:1}

\subsection{Derivation of forward Compton scattering sum rules}

The forward-scattering amplitude describing the elastic scattering
of a photon on a target with spin $s$ (real Compton scattering) is
characterized by $2s+1$ scalar functions which depend on a single
kinematic variable, {\it e.g.}, the photon energy $\nu$. In the
low-energy limit each of these functions corresponds to an
electromagnetic moment---charge, magnetic dipole, electric
quadrupole, {\it etc.}---of the target. In the case of a spin-1/2
target, such as the nucleon, 
the forward Compton amplitude is generally written as
\begin{eqnarray}
\label{DDeq2.2.2} T(\nu) = {\vec{\veps}\,'}^*\cdot\vec
\veps\,f(\nu)+ i\,\vec
\sigma\cdot({\vec{\veps}\,'}^*\times\vec\veps)\,g(\nu)\, ,
\end{eqnarray}
where $\vec \veps$, $\vec \veps\, '$ is the polarization vector of
the incident and scattered photon, respectively, while $\vec
\sigma$ are the Pauli matrices representing the dependence on the
target spin. The {\it
crossing symmetry} of the Compton amplitude of
Eq.~(\ref{DDeq2.2.2}) means invariance under
$\varepsilon'\leftrightarrow\varepsilon$, $\nu\leftrightarrow-\nu$, which
obviously leads to $f(\nu)$ being an even and
$g(\nu)$ being an odd function of the energy~: 
$f(\nu) = f(-\nu)$, $g(\nu) =-g(-\nu)$.
The two scalar functions $f(\nu),\,
g(\nu)$ admit the following low-energy expansion,
\begin{eqnarray}
f(\nu) & = & -\frac{e^2}{4\pi M} + (\alpha_E+\beta_M)\,\nu^2
+ {\mathcal{O}}(\nu^4) \ , 
\label{DDeq2.2.12} \\
g(\nu) & = & -\frac{e^2\kappa^2}{8\pi M^2}\,\nu +
\gamma_{0}\nu^3 + {\mathcal{O}}(\nu^5) \ , 
\label{DDeq2.2.13}
\end{eqnarray}
and hence, in the low-energy limit, are given in terms of the
target's charge $e$, mass $M$,  and anomalous magnetic moment (a.m.m.)
$\kappa$. The next-to-leading order terms are given in terms of
the nucleon electric ($\al_E$), magnetic ($\be_M$), and forward
spin ($\ga_0$) polarizabilities.

In order to derive sum rules (SRs) for these quantities one
assumes the scattering amplitude is an {\it analytic} function of
$\nu$ everywhere but the real axis, which allows writing
the real parts of the functions  $f(\nu)$ and $g(\nu)$ as a {\it
dispersion integral} involving their corresponding imaginary
parts. The latter, on the other hand, can be related to
combinations of doubly polarized photoabsorption cross-sections
via the {\it optical theorem},
\begin{eqnarray}
\mbox{Im}\ f(\nu) & = & \frac{\nu}{8\pi}
\left[\sigma_{1/2}(\nu)+\sigma_{3/2}(\nu)\right] \,, \\
\mbox{Im}\ g(\nu) & = & \frac{\nu}{8\pi}
\left[\sigma_{1/2}(\nu)-\sigma_{3/2}(\nu)\right] \,,
\label{DDeq2.2.6}
\end{eqnarray}
where $\sigma_{\la}$ is the doubly-polarized total cross-section
of the photoabsorption processes, with $\lambda$ specifying the
total helicity of the initial  system.  Averaging over the
polarization of initial particles gives the total unpolarized
cross-section,
 $\sigma_T=\half ( \sigma_{1/2}+\sigma_{3/2})$.

After these steps one arrives at the results (see, {\it e.g.},
\cite{Drechsel:2002ar} for more details): 
\bea 
\label{eq:drf}
f(\nu) & = & f(0) \,+\, \frac{\nu^2}{2\pi^2}\,
\int_{0}^{\infty}\frac{\sigma_T(\nu')}
{\nu'^2-\nu^2 - i\eps} \, d\nu'\ , \\
\label{eq:drg}
 g(\nu) &=& -\frac{\nu}{4\pi^2}\,
\int_{0}^{\infty}\frac{\De\sigma(\nu')} {\nu'^2-\nu^2 -
i\eps}\,\nu'\,d\nu'\ , \eea with $\De\sigma\equiv
\sigma_{3/2}-\sigma_{1/2}$, and where the sum rule for the
unpolarized forward amplitude $f(\nu)$ has been once-subtracted
to guarantee convergence. These relations can then be expanded in
energy to obtain the SRs for the different static properties
introduced in Eqs.~(\ref{DDeq2.2.12},\ref{DDeq2.2.13}). 
In this way we obtain the Baldin SR~\cite{Bal60,Lap63}:
\begin{eqnarray}
\label{eq:baldinsr}
\alpha_E + \beta_M = \frac{1}{2\pi^2}\,
\int_{0}^{\infty}\,\frac{\sigma_T(\nu)}{\nu^2}
\,d\nu \ ,
\end{eqnarray}
the GDH SR:
\begin{eqnarray}
\label{eq:gdhsr}
\frac{e^2\kappa^2}{2M^2}=\frac{1}{\pi}
\int_{0}^{\infty}\,\frac{\De\sigma(\nu)}{\nu}\,d\nu \, ,
\end{eqnarray}
a SR for the forward spin polarizability:
\begin{eqnarray}
\label{eq:gamma0sr}
\gamma_0= \,-\,\frac{1}{4\pi^2}\,\int_{0}^{\infty}\,
\frac{\De\sigma(\nu)}
{\nu^3}\,d\nu\ ,
\end{eqnarray}
and, in principle, one could continue in order to isolate higher
order moments \cite{Holstein:1999uu}.

\begin{figure}
\resizebox{0.45\textwidth}{!}{%
  \includegraphics{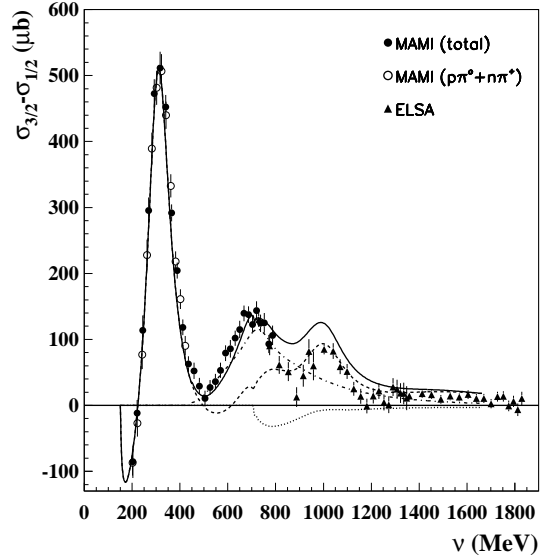}
}
\caption{The helicity difference
$\sigma_{3/2}(\nu)-\sigma_{1/2}(\nu)$ for the proton.
The calculations include the contribution of $\pi N$ intermediate
states (dashed curve) \cite{Drechsel:1998hk}, $\eta N$ intermediate state
(dotted curve) \cite{Dre01}, and the $\pi\pi N$ intermediate
states
(dashed-dotted curve) \cite{Hol01}. The total sum of these
contributions is shown by the full curves. The MAMI data are
from Ref.~\cite{Ahr00,Ahr01} and the ELSA data from 
Ref.~\cite{Dutz:2003mm}.
}
\label{DDfig2.2.3}
\end{figure}

Recently, the helicity difference $\Delta \sigma$ which enters the integrands 
 of Eqs.~(\ref{eq:gdhsr}) and (\ref{eq:gamma0sr}) has been measured. 
The first measurement was carried out at MAMI (Mainz) for photon
energies in the range 200~MeV$ < \nu < 800~$MeV~\cite{Ahr00,Ahr01}, 
and was extended at ELSA (Bonn)~\cite{Dutz:2003mm}  
for $\nu$ up to 3~GeV. 
This difference, shown in Fig.~\ref{DDfig2.2.3}, fluctuates much
more strongly than the total cross section $\sigma_T$. The
threshold region is dominated by S-wave pion production, 
and therefore mostly
contributes to the cross section $\sigma_{1/2}$. In the region of
the $\Delta (1232)$ with spin $J=3/2$, both helicity cross
sections contribute, but since the transition is essentially $M1$,
we find $\sigma_{3/2}/\sigma_{1/2}\approx3$. 
As seen from Fig.~\ref{DDfig2.2.3}, $\sigma_{3/2}$ also 
dominates the proton photoabsorption cross section
in the second and third resonance regions. 

\subsection{Linearized GDH sum rule}

Recently, it was shown that by taking
derivatives of the GDH sum rule with respect to the \amm\ 
one can obtain a new set of sum rule-like relations with intriguing 
properties~\cite{Pascalutsa:2004ga,Holstein:2005db}.  

To derive such sum rules. one 
begins by introducing a `classical' (or `trial') value of the
particle's \amm, $\kappa_0$. At the Lagrangian level this amounts to
the introduction of a Pauli term for the spin-1/2 field~: \beq
{\cal L}_{\mbox{\small Pauli}}= \frac{i\kappa_0}{4M} \, \bar
\psi\, \si_{\mu\nu}\, \psi\, F^{\mu\nu}\,, \eeq where $F^{\mu
\nu}$ is the electromagnetic field tensor and $\si_{\mu\nu}=(i/2)
[\ga_\mu,\ga_\nu]$ is the usual Dirac tensor operator.  At the end
of the calculation, $\kappa_0$ is set to zero, but in the 
evaluation of the absorption cross sections 
the total value of the \amm\ is $\kappa=\kappa_0 +\delta \kappa$,
with $\de\kappa$ denoting the loop contribution.  
It was shown in Ref.~\cite{Pascalutsa:2004ga,Holstein:2005db} 
that this yields the SR~: 
\beq
\label{linsr} \frac{4\pi^2\al_{em}}{M^2}\, \kappa  =
\int_{0}^\infty \! \left. \De\si'(\nu)\right|_{\kappa_0=0}\,
\frac{d\nu}{\nu}\,,  
\eeq 
where  $\De\si'(\nu)$ is the derivative of an absorption 
cross section w.r.t. the trial \amm\ value $\kappa_0$. 
The striking feature of this sum rule is
the {\it linear} relation between the \amm\ and the (derivative of
the) photoabsorption cross section, in contrast to the GDH SR
where $\kappa$ appears quadratically. Although the
cross-section quantity $\De\si'(\nu)$ is not an observable, it
is very clear how it can be determined within a specific theory.
Thus, for example, the first derivative of the tree-level
cross-section with respect to $\kappa_0$, at $\kappa_0=0$, in QED
was worked out in Ref.~\cite{Pascalutsa:2004ga},  
yielding Schwinger's one-loop result. 
It is noteworthy that this result is reproduced 
by computing only a (derivative of the) {\it tree-level} Compton
scattering cross-section and then performing an integration over
energy. This is definitely much simpler than obtaining the
Schwinger result from the GDH SR directly~\cite{DiV01}, which
requires an input at the one-loop level. 

\begin{figure}
\resizebox{0.5\textwidth}{!}{%
  \includegraphics{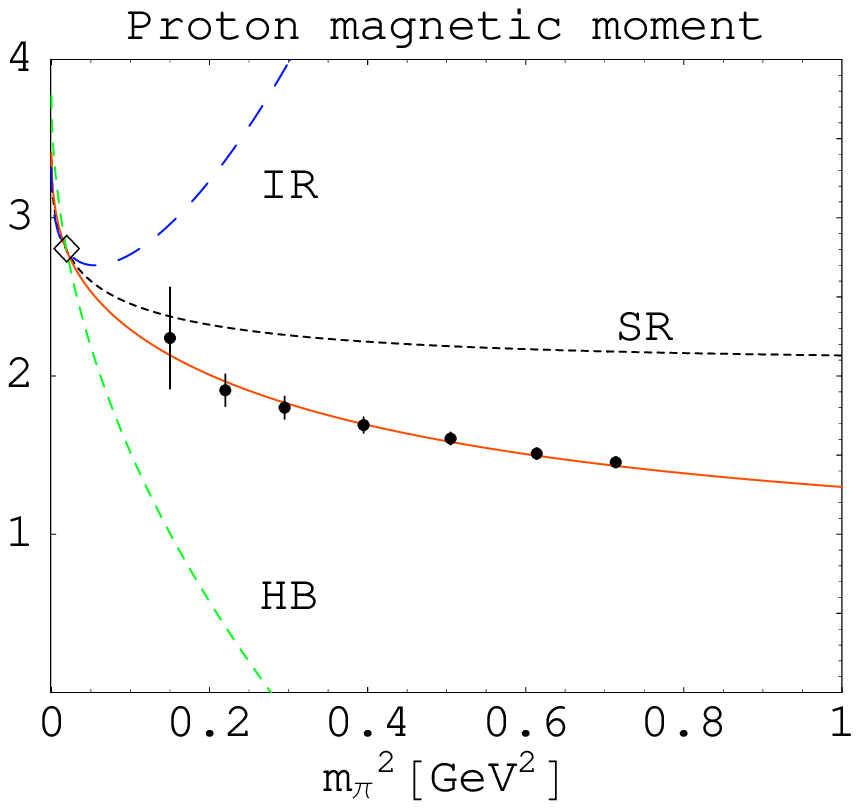}
  \includegraphics{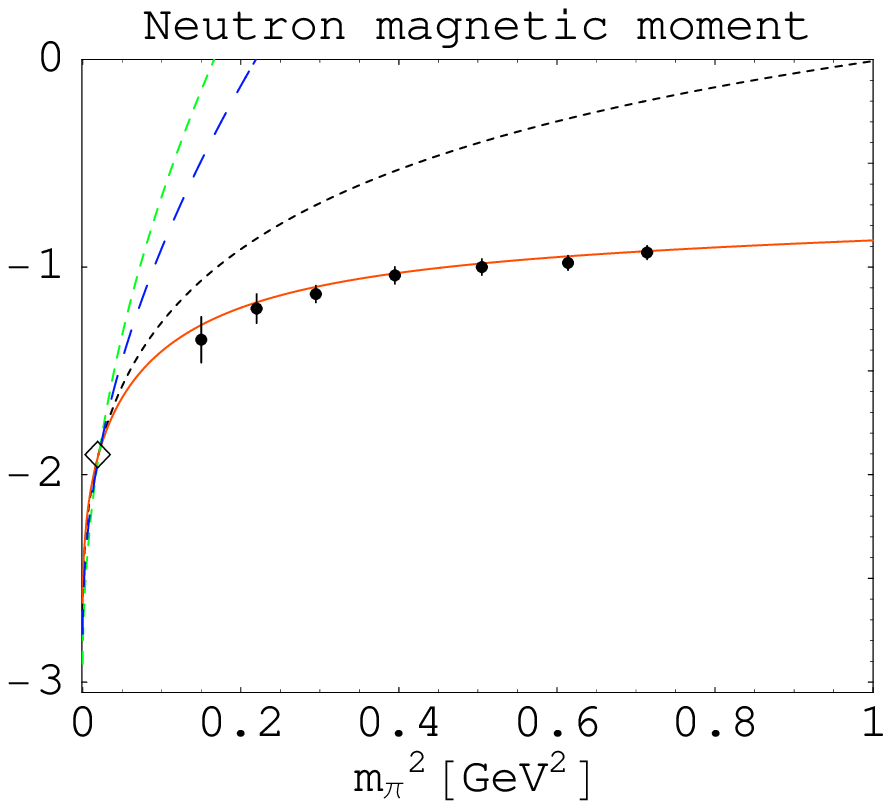}
}
\caption{ Chiral behavior of proton and neutron magnetic
moments (in nucleon magnetons) to one loop compared with lattice
data (solid circles). 
``SR'' (dotted lines): one-loop relativistic result based on 
Eq.~(\ref{linsr}),
``IR'' (blue long-dashed lines): infrared-regularized relativistic
result, ``HB'' (green dashed lines): leading non-analytic 
term in the heavy-baryon expansion. 
Red solid lines:  single-parameter  fit based on the SR
result, see Refs.~\cite{Pascalutsa:2004ga,Holstein:2005db}. 
The open
diamonds represent the experimental values at the physical pion
mass.
} 
\label{chibehavior}
\end{figure}

The SR of Eq.~(\ref{linsr}) can furthermore be applied to 
study the magnetic moment and polarizabilities of the nucleon
in a relativistic chiral 
EFT framework~\cite{Pascalutsa:2004ga,Holstein:2005db}. In 
particular it allows to study the chiral extrapolation of these quantities, 
as shown in Fig.~\ref{chibehavior} for the magnetic moments.
One sees that the SR calculation, 
strictly satisfying analyticity, is better suited 
for the chiral extrapolation of lattice QCD results than the usual 
heavy-baryon expansions or the ``infrared-regularized'' relativistic 
theory. 
 
\section{Nucleon excitation spectrum} 
\label{sec:2}

The sum rules for Compton scattering off the nucleon are dominated by 
its first excited state --- the $\Delta(1232)$ resonance, 
as is apparent from Fig.~\ref{DDfig2.2.3}.   
Through the sum rules,  
the $\Delta$ therefore plays a preponderant role in our understanding 
of low-energy nucleon structure. This justifies a dedicated 
effort to study this resonance. 

High-precision measurements of the $N$-to-$\De$
transition by means of electromagnetic probes became possible with the
advent of the new generation of electron scattering facilities, such as 
BATES, MAMI, and JLab,  many measurements being completed
in recent years~\cite{Beck:1999ge,LEGS97,Bates01,Jlab}.

The {\it electromagnetic}  nucleon-to-$\De$ (or, in short $\ga N \De$)  
transition is predominantly
of the magnetic dipole ($M1$) type. In a simple quark-model picture, 
this $M1$ transition 
is described by a spin flip of a quark in the $s$-wave state. 
Any $d$-wave admixture
in the nucleon {\it or} the $\Delta$ wave-functions allows for the electric ($E2$) and Coulomb ($C2$)
quadrupole transitions. Therefore by measuring these one is able to assess the presence
of the $d$-wave components and hence quantify to which extent the nucleon or the $\De$ wave-function
deviates from the spherical shape, {\it i.e.}, to which extent they are 
``deformed''~\cite{deformation}. 
The $\ga N\Delta$ transition, on the other hand,  was
accurately measured in the pion photo- and electro-production reactions 
in the $\De$-resonance energy region. The $E2$ and $C2$ transitions 
 were found to be relatively small at moderate  momentum-transfers ($Q^2$), 
the ratios $R_{EM}=E2/M1$ and 
$R_{SM}=C2/M1 $ are at the level of a few percent.  

Traditionally, the resonance parameters are extracted 
by using {\it unitary isobar models}~\cite{ELAs,SpainModel,MV95,KVItheory,GiessenModel,Drechsel:1998hk,Inna}, which
in essence are unitarized tree-level calculations based on
phenomenological Lagrangians. However, 
at low  $Q^2$ the $\gamma N \Delta$-transition 
shows great sensitivity to the ``pion 
cloud'', which until recently could only be
comprehensively studied within 
{\it dynamical models}~\cite{Sato2,GrS96,DMT,Fuda03,PaT00,Caia04}, 
which - unlike the isobar models - include quantum effects
due to pion loops.

With the advent of the chiral effective field theory (\ceft) 
of QCD~\cite{Weinberg:1978kz,Gasser:1983yg}
and its extensions to the $\De$(1232) 
region~\cite{SSE1,SSE2,Tang96,Fettes01,PP03,Hacker05}, 
it has become possible to study the nucleon and $\Delta$-resonance properties 
in a profoundly different way. Recently,
first relativistic \ceft\  studies were performed of the 
$\gamma N \Delta$-transition 
in pion electroproduction~\cite{Pascalutsa:2005ts,PV06} and 
of the $\De$(1232) magnetic dipole moment (MDM) in the radiative
pion photoproduction~\cite{PV05}. The advantages over the
previous dynamical approaches are apparent: \ceft\ is
a low-energy effective field theory of QCD and as such 
it provides a firm theoretical foundation, with all
the relevant symmetries and scales of QCD built in consistently.

The \ceft\ of the strong interaction
is indispensable, at least at present, in relating the low-energy observables
(e.g., hadron masses, magnetic moments, form factors)  
to {\it ab initio} QCD calculations on the lattice. 
On the other hand, \ceft\ can and should be used 
in extracting various hadronic properties
from the experiment. The \ceft\ fulfills both 
of these roles in a gratifying fashion.

The following sections review recent progress in 
the \ceft\ in the $\Delta$-resonance region that has been obtained  
for the $N$ and $\Delta$ masses~\cite{PV05self}, 
the $\gamma N \Delta$ transition~\cite{Pascalutsa:2005ts,PV06},
and the $\De$ MDM~\cite{PV05}.

\subsection{Chiral effective field theory in the $\Delta(1232)$ region}

Starting from the effective Lagrangian of 
chiral perturbation theory ($\chi$PT) with pion and nucleon
fields~\cite{GSS89}, 
the $\De$ is included explicitly in the so-called $\de$-expansion
scheme~\cite{PP03}. 
In the following, the Lagrangian $\lag^{(i)}$ is organized  
such that superscript $i$ stands for 
the power of electromagnetic coupling $e$ plus the number of derivatives
of pion and photon fields. Writing here only the terms involving the 
spin-3/2 isospin-3/2 field $\De^\mu$ of the $\De$-isobar 
gives:\footnote{Here we introduce totally antisymmetric products
of $\ga$-matrices:
$\ga^{\mu\nu}=\half[\ga^\mu,\ga^\nu]$,
$\ga^{\mu\nu\al}=\half \{\ga^{\mu\nu},\ga^\al\}=
i\veps^{\mu\nu\al\be}\ga_\be\ga_5$. }
\bea
\lag^{(1)}_{N\De} &=&  \ol \De_\mu \left(i\ga^{\mu\nu\al}\,D_\al - 
M_\De\,\ga^{\mu\nu}\right) \De_\nu \nn\\
&+& \!\frac{i h_A}{2 f_\pi M_\De}\left\{
\ol N\, T_a \,\ga^{\mu\nu\la}\, (\pa_\mu \De_\nu)\, D_\la \pi^a 
+ \mbox{H.c.}\right\} \nn \\
&-& \frac{H_A}{2M_\De f_\pi}  \,
\veps^{\mu\nu\rho\si} \,\ol\De_\mu 
\, \Tau^a \,(\pa_\rho \De_\nu )\,\pa_\si \pi^{a}, 
\label{lagran1} \\
\lag^{(2)}_{N\De} &=&  \frac{i e (\mu_\De-1)}{2M_\De}\, \ol \De_\mu \De_\nu\, F^{\mu\nu} 
\nn\\
&+& \frac{3 i e g_M}{2M_N (M_N+M_\Delta)}\left\{\ol N\, T_3
\,\pa_{\mu}\De_\nu \, \tilde F^{\mu\nu}+ \mbox{H.c.}\right\} \nn\\
&-& \!\frac{e h_A}{2 f_\pi M_\De}\left\{
\ol N\, T_a\,\ga^{\mu\nu\la} A_\mu \De_\nu\, \pa_\la \pi^a 
+ \mbox{H.c.}\right\}, 
\label{lagran2} \\
\lag^{(3)}_{N\De} &=&  \frac{-3 e}{2M_N (M_N + M_\Delta)} \ol N \, T^3
\ga_5 \left[ g_E (\pa_{\mu}\De_\nu) 
\right. \nn \\
&+& \left. \frac{i g_C}{M_\De} \ga^\al  
(\pa_{\al}\De_\nu-\pa_\nu\De_\al) \,\pa_\mu\right] F^{\mu\nu}+ \mbox{H.c.}, 
\label{lagran3}
\eea
where $M_N$ and $M_\De$ are, respectively, the nucleon and $\De$-isobar masses,
$N$ and $\pi^a\,\, (a=1,2,3)$ stand for the nucleon and pion fields, $D_\mu$ is the covariant 
derivative ensuring the electromagnetic gauge-invariance, $F^{\mu\nu}$ and $\tilde F^{\mu\nu}$
are the electromagnetic field strength and its dual,  
$T_a$ are the isospin 1/2 to 3/2 transition matrices, 
and $\Tau^a$ are the generators in the isospin 3/2 representation 
of SU(2), satisfying $\Tau^a \Tau^a = 5/3$. 
The coupling constants are given by~: $f_\pi = 92.4$ MeV, 
$h_A\simeq 2.85$ is obtained from 
the $\De$-resonance width, $\Gamma_\Delta  = 0.115$~GeV,   
and for $H_A$ the large-$N_c$ relation $H_A=(9/5) g_A$ is adopted, 
with $g_A\simeq 1.267$ the nucleon axial-coupling constant.
 
Note that the electric and the Coulomb $\ga N\De$ couplings 
($g_E$ and $g_C$ respectively) are of one order higher than
the magnetic ($g_M$) one, because of the $\ga_5$ which involves 
the ``small components'' of the fermion fields and thus 
introduces an extra power of the 3-momentum.
The MDM $\mu_\De$ is defined here in units of $[e/2M_\De]$. 
Higher electromagnetic moments are omitted, because they do not contribute 
at the orders that we consider.

Note that $\lag^{(1)}_\De$ contains the free Lagrangian, 
which is formulated in~\cite{RaS41} such that the number of 
spin degrees of freedom of the relativistic spin-3/2 field is constrained
to the physical number: $2s+1=4$.  The $N$ to $\De$ transition 
couplings in Eqs.~(\ref{lagran1},\ref{lagran2},\ref{lagran3})
are consistent with these 
constraints~\cite{Pas98,Pascalutsa:1999zz,Pas01}. 
The  $\ga \De\De$ coupling is more subtle since in this case
constraints do not hold for sufficiently strong electromagnetic fields, 
see, e.g., \cite{DPW00}. 
In extracting the $\De$ MDM, 
it is therefore assumed that the electromagnetic 
field is weak, compared to the $\Delta$ mass scale.  

The inclusion of the 
$\De$-resonance introduces another light scale - besides the pion mass - 
in the theory,
the resonance excitation energy: $\De\equiv M_\De-M_N\sim 0.3$ GeV.
This energy scale is still relatively light in comparison to
the chiral symmetry breaking scale
$\La_{\chi SB} \sim 1$ GeV. Therefore,
$\de = \De/\La_{\chi SB}$ can be treated as a small parameter. 
The question is, how to compare this parameter
with the small parameter of chiral perturbation theory ($\chi$PT),
$\eps = m_\pi /\La_{\chi SB}$. 
 
In most of the literature 
(see, e.g.,Refs.~\cite{SSE1,SSE2,Tang96,Fettes01,Hacker05}) 
they are assumed to be of 
comparable size, $\de\approx \eps$. 
This, however, leads to a somewhat unsatisfactory
result because obviously the $\De$-contributions
are {\it overestimated} at lower energies and {\it underestimated}
at the resonance energies. To estimate the $\De$-resonance
contributions correctly, and depending on the energy region,
one needs to count $\de$ and $\eps$ {\it differently}.

A relation $\eps =\de^2$ was suggested and explored in~\cite{PP03},
and is referred to as the $\delta$-expansion. 
The second power is indeed the closest integer power for
the relation of these parameters in the real world.
In Refs.~\cite{PV05self,Pascalutsa:2005ts,PV06,PV05}  
this relation was used for power-counting purposes
only, and was not imposed in the actual evaluations of diagrams.
Each diagram is simply  characterized 
by an overall $\delta$-counting index $n$,
which tells us that its contribution begins at O($\de^n$). 

Because of the distinction of $m_\pi$ and $\De$ 
the counting of a given diagram depends 
on whether the characteristic momentum $p$ is  
in the low-energy region ($p\sim m_\pi$) or in the resonance
region ($p\sim \De$). 
In the low-energy region the index of a graph with $L$ loops, 
$N_\pi$ pion propagators, $N_N$ nucleon propagators, 
$N_{\De}$ $\Delta$-propagators, and $V_i$ vertices of dimension $i$ is 
\beq
n= 
2 \,(\,\sum_i i V_i + 4 L  - N_N - 2 N_\pi ) 
-N_\De \equiv 2 n_{\chi{\mathrm PT}} - N_\De, 
\eeq
where $ n_{\chi{\mathrm PT}} $
is the index in $\chi$PT with no $\De$'s \cite{GSS89}.
In the resonance region, one distinguishes the one-$\De$-reducible (O$\De$R) graphs~\cite{PP03}. Such graphs contain $\De$ propagators
which go as $1/(p-\De)$, and hence for $p\sim \De$ they are large and 
all need to be included. This gives an incentive, {\it within
the power-counting scheme}, to 
resum $\De$ contributions.
Their resummation amounts to dressing the $\De$ propagators 
so that they behave as $1/(p-\De-\Si)$. The self-energy 
$\Si$ begins at order $p^3$ and thus
a dressed O$\De$R propagator counts as $1/\de^3$.
If the number of
such propagators in a graph is $N_{O\De R}$, the power-counting index of
this graph in the resonance region is given by 
\beq
n=n_{\chi{\mathrm PT}} - N_\De - 2N_{O\De R},
\eeq 
where $N_\De$ is the total number
of $\De$-propagators.

A word on the renormalization program, as it is an indivisible part
of power counting in a relativistic theory. Indeed, without some
kind of renormalization the loop graphs diverge as $\La^{\cal N} $,
where $\La$ is an ultraviolet cutoff, and ${\cal N}$ is a positive
power proportional to the power-counting index of the graph. 
Also, contributions of heavy scales,
such as baryon masses, may appear as $M^{\cal N}$. 
The renormalization
of the loop graphs can and should be performed so as to absorb
these large contributions into the available low-energy constants,
thus bringing the result in accordance
with power counting \cite{Gegelia:1999gf}.

To give an example, consider the one-$\pi N$-loop contribution to
the nucleon mass. For the $\pi NN$ vertex, 
the power counting tells us that this contribution
begins at $O(m_\pi^3)$. An explicit calculation, however, will 
show (e.g., \cite{GSS89}) that the loop produces
$O(m_\pi^0)$ and $O(m_\pi^2)$ terms, both of which are (infinitely) large.
{\it This is not a violation of power counting}, because there are two
low-energy constants: the nucleon mass in the chiral limit, $M^{(0)}$,
and $c_{1N}$, which enter at order  $O(m_\pi^0)$ and $O(m_\pi^2)$, 
respectively, and {\it renormalize away} 
the large contributions coming from the loop.    
The renormalized relativistic result, up to and including
$O(m_\pi^3)$,  can be written as~\cite{PV05self}:
\begin{eqnarray}
M_N &=& M_N^{(0)} - 4 \, c_{1 N} \, m_\pi^2  
\label{eq:nucpin} \\ 
&-& \frac{3 \, g_A^2}{(8 \pi f_\pi)^2} \, m_\pi^3 \, 
\left\{ 4  \left( 1 - \frac{m_\pi^2}{4 M_N^2} \right)^{5/2} 
\arccos\frac{m_\pi}{2 M_N} \right. \nn \\
&&+ \frac{17 m_\pi}{16 M_N}  
- \left(\frac{m_\pi}{2 M_N}\right)^3  \nonumber \\ 
&&\left. + \, \frac{m_\pi}{8 M_N} 
\left[ 30 - 10 \left(\frac{m_\pi}{M_N}\right)^2 
+  \left(\frac{m_\pi}{M_N}\right)^4 \right] 
\, \ln \frac{m_\pi}{M_N}  
\right\} , \nn
\end{eqnarray}
and one can easily verify that the loop contribution begins
at $O(m_\pi^3)$ in agreement with power counting. 

Likewise, the $\Delta$ mass has also been calculated in relativistic \ceft\, 
see Ref.~\cite{PV05self} for details. 

\begin{figure}
\resizebox{0.5\textwidth}{!}{%
  \includegraphics{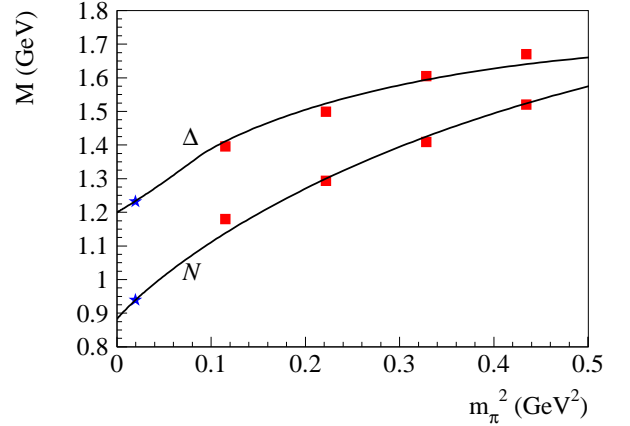}
}
\caption{Pion-mass dependence of the 
nucleon and $\Delta(1232)$ masses. 
The curves are two-parameter expressions for the $\pi N$ loop contributions 
to $M_N$ and $M_\Delta$ as calculated in Ref.~\cite{PV05self} 
(see text for the values of the low-energy constants).  
The red squares are lattice results from the MILC 
Collaboration~\protect\cite{Bernard:2001av}. 
The stars represent the physical mass values. }
\label{nucdelmass}
\end{figure}

The $m_\pi$ dependence of the nucleon and $\De$-resonance
masses are compared with lattice results in 
Fig.~\ref{nucdelmass}. One of the two parameters
in Eq.~(\ref{eq:nucpin}) is constrained  
by the physical nucleon mass value at $m_\pi = 0.139$ GeV,  
while the other parameter
is fit to the lattice data shown in the figure. 
This yields~: $M_N^{(0)} = 0.883$~GeV and $c_{1 N} = -0.87$~GeV$^{-1}$. 
As is seen from the figure, with this two-parameter form for $M_N$, 
a good description of lattice results is obtained 
up to $m_\pi^2 \simeq 0.5$~GeV$^2$. 
Analogously to the nucleon case, 
one low-energy constant for the $\Delta$ is fixed 
from the physical value of the $\Delta$ mass,
while the second parameter is fit to the lattice data
shown in Fig.~\ref{nucdelmass},
yielding~: $M_\Delta^{(0)} = 1.20$~GeV and 
$c_{1 \Delta} = -0.40$~GeV$^{-1}$. 
As well as for the nucleon, 
this two-parameter form for $M_\Delta$ yields a fairly good description 
of the lattice results up to $m_\pi^2 \simeq 0.5$~GeV$^2$.

\subsection{$\gamma N \Delta$ transition}

\begin{figure}
\resizebox{0.45\textwidth}{!}{%
  \includegraphics{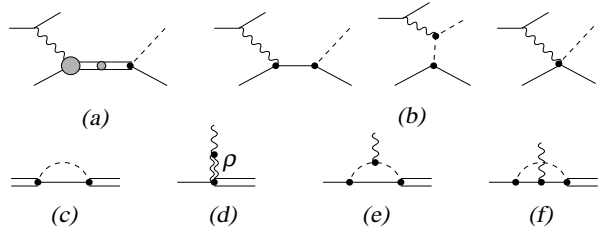} 
}
\caption{Diagrams for the $e N \to e \pi N $ reaction 
at NLO in the $\delta$-expansion. 
Double lines represent the $\De$ propagators.
}
\label{diagramsgandel}
\end{figure}

The $\gamma N \Delta$ transition is usually studied through the 
pion electroproduction process.
The pion electroproduction amplitude to NLO in the $\de$ expansion, in the
resonance region, is given by graphs in Fig.~\ref{diagramsgandel}(a) and (b), 
where the shaded blobs in graph (a) include  corrections 
depicted in Fig.~\ref{diagramsgandel}(c--f). The hadronic part of
graph (a) begins at ${\cal O}(\de^0)$ which here is the leading order. 
The Born graphs (b) contribute at ${\cal O}(\de)$.
The one-loop vertex corrections of Fig.~\ref{diagramsgandel}(e) and (f)  
to the $\gamma N \Delta$-transition form factors 
have been evaluated in two independent ways in 
Refs.~\cite{Pascalutsa:2005ts,PV06}, to which we refer for details.  
At NLO there are also vertex corrections 
of the type (e) and (f) with nucleon propagators in the loop
replaced by the $\De$-propagators. However, after the appropriate
renormalizations and $Q^2\ll\La_{\chi SB}\De$, 
these graphs start to contribute at next-next-to-leading order.
The vector-meson diagram, Fig.~\ref{diagramsgandel}(d), 
contributes to NLO for $Q^2\sim \La_{\chi SB} \De$. It was included 
effectively in Refs.~\cite{Pascalutsa:2005ts,PV06} 
by giving the $g_M$-term a dipole $Q^2$-dependence, 
in analogy to how it is usually done for the nucleon isovector form factor.

\begin{figure}
\resizebox{0.35\textwidth}{!}{%
\hspace{1cm}
  \includegraphics{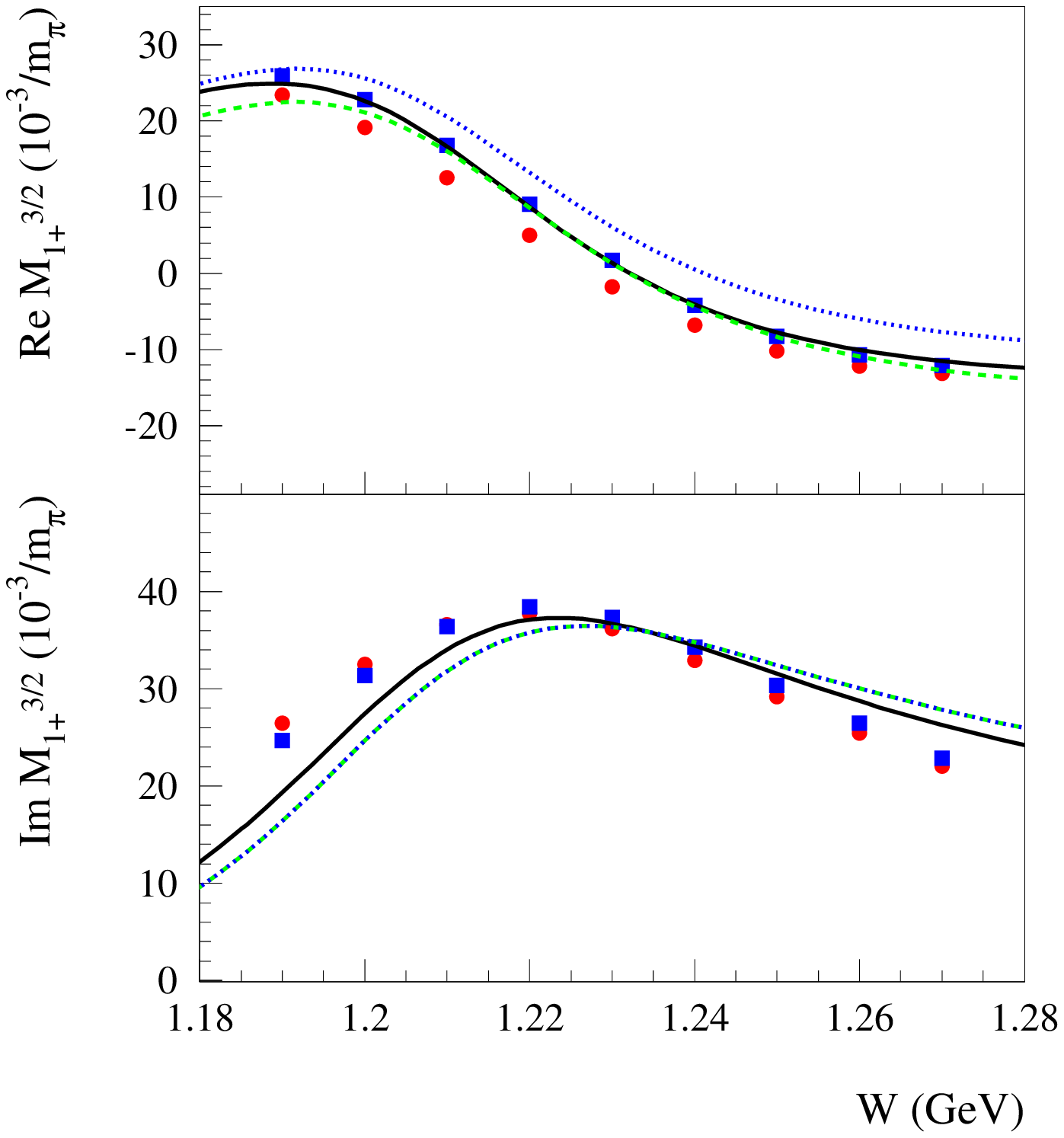}
}
\resizebox{0.35\textwidth}{!}{%
\hspace{1cm}
  \includegraphics{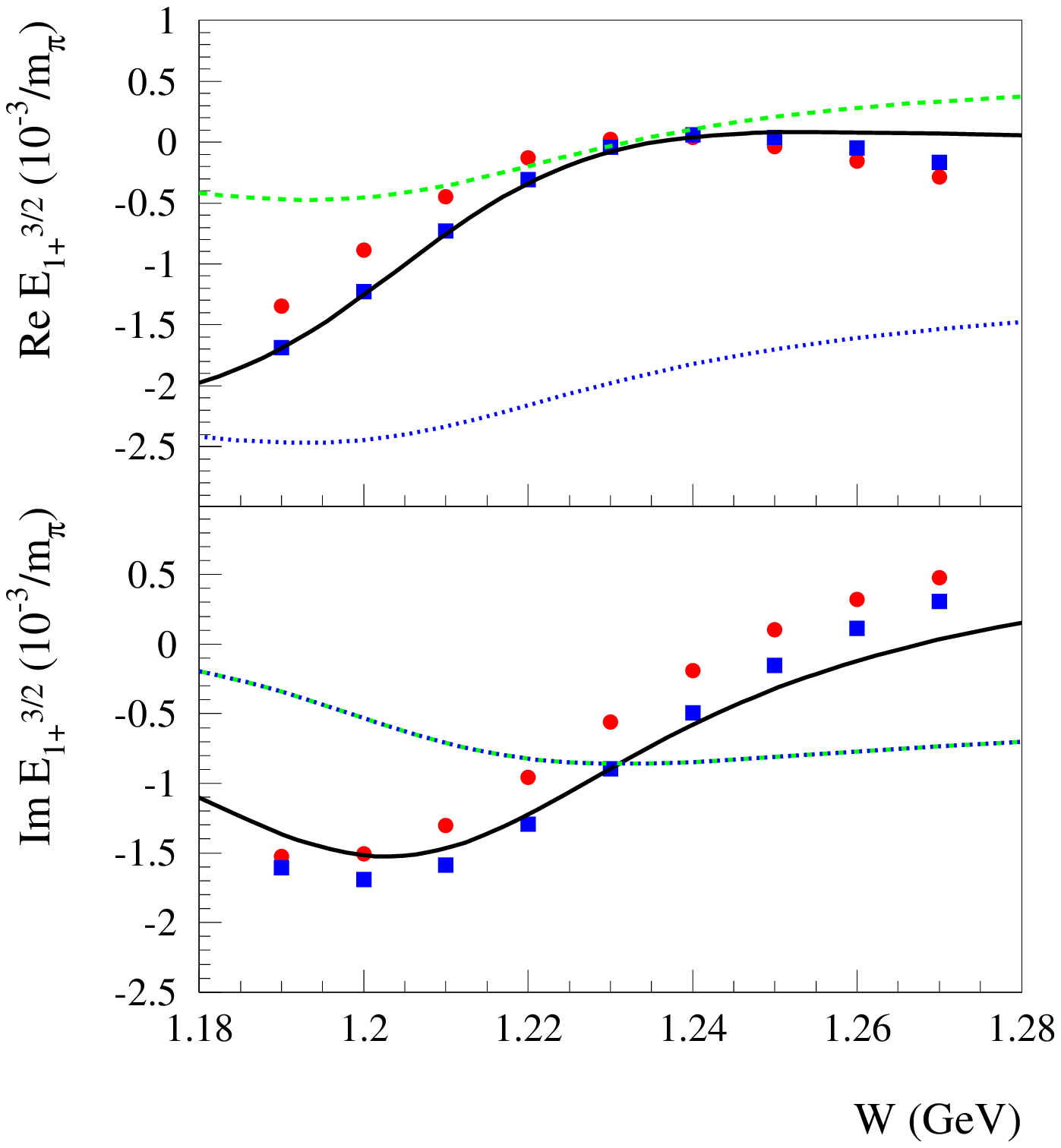}
}
\caption{
Multipole amplitudes
$M_{1+}^{(3/2)}$ (top panels) and $E_{1+}^{(3/2)}$ (bottom panels)
for pion photoproduction as function of the invariant mass $W$ of 
the $\pi N$ system.  
Dashed curves: 
$\Delta$ contribution without the $\ga N\De$-vertex corrections,
[i.e., Fig.~\ref{diagramsgandel}(a) 
without  Fig.~\ref{diagramsgandel}(e, f)].  
Dotted curves: adding the Born contributions, 
Fig.~\ref{diagramsgandel}(b),
 to the dashed curves. 
Solid curves: complete NLO calculation, includes 
all graphs from Fig.~\ref{diagramsgandel}. In all curves
the low-energy parameters are chosen as~: 
$g_M = 2.9$, $g_E = -1.0$. 
The data point are from the 
SAID analysis~(FA04K)~\protect\cite{Arndt:2002xv} (red circles), and from the 
MAID 2003 analysis~\protect\cite{Drechsel:1998hk} (blue squares).
}
\label{fig:gap_pin_mult}
\end{figure}

The resonant pion photoproduction multipoles are used 
to determine the two low-energy constants:
$g_M$ and $g_E$, the strength of the $M1$ and $E2$ 
$\gamma N \Delta$ transitions. 
In Fig.~\ref{fig:gap_pin_mult}, we show the result of the \ceft\  calculations 
for the pion photoproduction resonant multipoles 
$M_{1+}^{(3/2)}$ and $E_{1+}^{(3/2)}$, around the 
resonance position, as function of the total {\it c.m.} energy $W$ 
of the $\pi N$ system. 
These two multipoles are well established by the MAID~\cite{Drechsel:1998hk}
and SAID~\cite{Arndt:2002xv} partial-wave solutions which allow us to fit 
the two low-energy constants of the chiral Lagrangian 
of Eqs.~(\ref{lagran2},\ref{lagran3}) 
as~:~$g_M = 2.9$, $g_E = -1.0$.   As is seen from Fig.~\ref{fig:gap_pin_mult},
with these values the NLO results (solid lines) give a 
good description of the energy dependence of the resonant multipoles in 
a window of 100 MeV around the $\Delta$-resonance position.
Also, these values yield $R_{EM}= -2.3$ \%, 
in a nice agreement with experiment~\cite{Beck:1999ge}.

The dashed curves in Fig.~\ref{fig:gap_pin_mult} 
show the contribution of the $\Delta$-resonant diagram of 
Fig.~\ref{diagramsgandel}(a)
{\it without} the NLO vertex corrections Fig.~\ref{diagramsgandel}(e, f).
For the $M_{1+}$ multipole this is the LO contribution.
For the $E_{1+}$ multipole
the LO contribution is absent [the $g_E$ coupling
is of one order higher than $g_M$]. 
Hence,  the dashed curve represents
a partial NLO contribution to $E_{1+}$ therein.
Upon adding the non-resonant  Born graphs, Fig.~\ref{diagramsgandel}(b), to the
dashed curves one obtains the dotted curves in 
Fig.~\ref{fig:gap_pin_mult}. These non-resonant contributions are purely 
real at this order and do not affect the imaginary part of the multipoles. 
One sees that the resulting calculation is flawed because the real 
parts of the resonant multipoles now fail to cross zero at the resonance 
position and hence unitarity, in the sense of Watson's theorem~\cite{Wat54},
is violated.
The complete NLO calculation, shown by the solid curves in  
Fig.~\ref{fig:gap_pin_mult}, 
includes in addition the vertex 
corrections, Fig.~\ref{diagramsgandel}(e, f), which restore unitarity 
{\it exactly}. {\it Watson's theorem is satisfied exactly by
the NLO, up to-one-loop  amplitude} 
given the graphs in Fig.~\ref{diagramsgandel}.

\begin{figure}
\resizebox{0.45\textwidth}{!}{%
\hspace{.5cm}
  \includegraphics{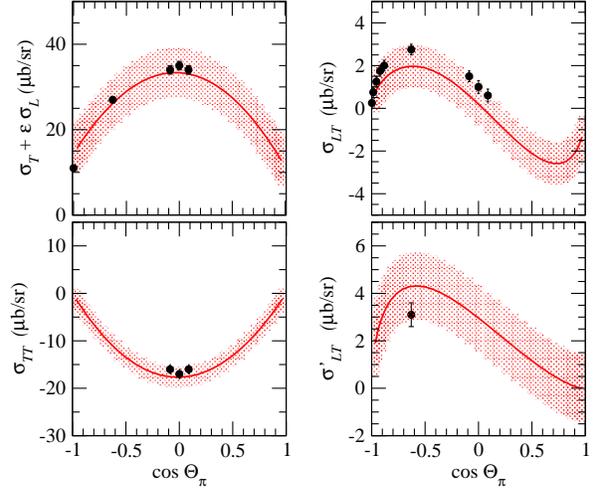}
}
\caption{\ceft\ NLO results for the  $\Th_\pi$ dependence of the 
$\ga^\ast p \to \pi^0 p$ cross sections 
at $W = 1.232$~GeV and $Q^2$ = 0.127~GeV$^2$. 
The theoretical error bands are  described in the text. 
Data points are from BATES experiments~\cite{Bates01,Kunz:2003we}.
}
\label{crossections}
\end{figure}

Fig.~\ref{crossections} shows the NLO results for  
different virtual photon absorption cross sections (for definitions, see 
Ref.~\cite{PV06}) at the resonance position, and 
for $Q^2 \simeq 0.127$~GeV$^2$, where recent precision data are available. 
Besides the low-energy constants $g_M$ and $g_E$, which were fixed 
from the resonant multipoles in Fig.~\ref{fig:gap_pin_mult}, 
the only other low-energy constant from Eq.~(\ref{lagran3}) 
entering the NLO electroproduction calculation is $g_C$. 
The main sensitivity on $g_C$ enters in 
$\sigma_{LT}$. A best description of the $\si_{LT}$ data in 
Fig.~\ref{crossections} is obtained by choosing 
$g_C = -2.36$.

The theoretical uncertainty due to the neglect of 
higher-order effects was estimated in Ref.~\cite{PV06}. 
We know that they must be suppressed
by at least one power of $\de$ ($=\De/\La_{\chi SB}$) 
as compared to the NLO and two powers of $\de$  
as compared to the LO contributions. 
These error estimates are shown by the bands in Fig.~\ref{crossections}.
One sees that the NLO \ceft\ calculation, within its accuracy,
is consistent with the experimental data for these observables.

\begin{figure}
\resizebox{0.4\textwidth}{!}{%
\hspace{.5cm}
  \includegraphics{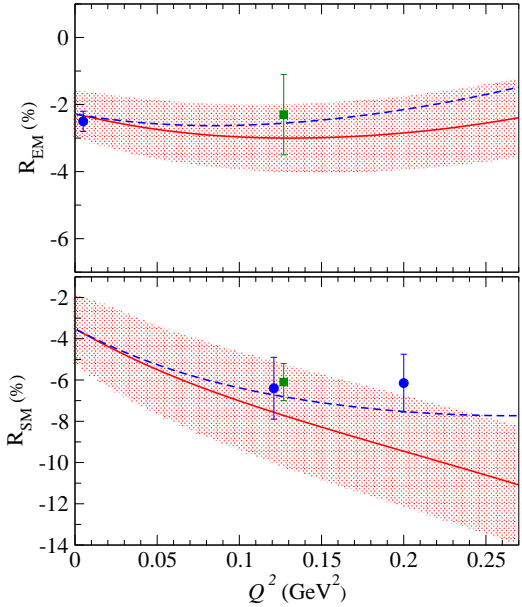}
}
\caption{
$Q^2$ dependence of the NLO results (solid curves) 
for $R_{EM}$ (upper panel) and  $R_{SM}$ 
(lower panel)~\cite{Pascalutsa:2005ts,PV06}. 
The blue dashed curves represent a phenomenological estimate 
of N$^2$LO effects by including $Q^2$-dependence 
in $g_E$ according to a dipole behavior, see Ref.~\cite{PV06}.   
The blue circles are data points from MAMI 
for $R_{EM}$~\protect\cite{Beck:1999ge}, and 
 $R_{SM}$~\protect\cite{Pospischil:2000ad,Elsner:2005cz}.  
The green squares are data points from BATES~\protect\cite{Bates01}.
}
\label{ratiosQ2}
\end{figure}

Fig.~\ref{ratiosQ2} shows 
the $Q^2$ dependence of the ratios $R_{EM}$ and $R_{SM}$. Having 
fixed the low energy constants $g_M$, $g_E$ and $g_C$, this $Q^2$ 
dependence follows as a prediction.  
The theoretical uncertainty here (shown by  
the error bands) was also estimated in Ref.~\cite{PV06} 
over the range  of $Q^2$ from 0 to 0.2~GeV$^2$. 
One sees that the NLO calculations are consistent 
with the experimental data for both of the ratios.

\begin{figure}
\resizebox{0.4\textwidth}{!}{%
\hspace{.5cm}
  \includegraphics{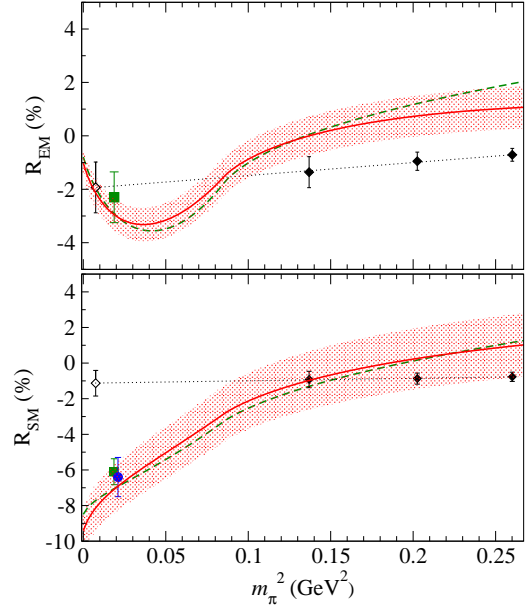}
}
\caption{
$m_\pi$ dependence of the NLO results at $Q^2=0.1$ GeV$^2$ for 
 $R_{EM}$ (upper panel) and 
$R_{SM}$ (lower panel).
The blue circle is a data point from MAMI~\protect\cite{Pospischil:2000ad}, 
the green squares are data points from BATES~\protect\cite{Bates01}. 
The three filled black diamonds at larger $m_\pi$   
are lattice calculations~\protect\cite{Ale05}, 
whereas the open diamond near $m_\pi \simeq 0$  
represents their extrapolation assuming linear dependence in $m_\pi^2$. 
Solid curves: NLO result when accounting for the $m_\pi$ dependence in 
$M_N$ and $M_\Delta$; 
Dashed curves: NLO 
result of Ref.~\protect\cite{Pascalutsa:2005ts}, where   
the $m_\pi$-dependence of $M_N$ and $M_\Delta$ was not accounted for. 
}
\label{ratios}
\end{figure}

Fig.~\ref{ratios} shows the $m_\pi$-dependence of the ratios 
$R_{EM}$ and $R_{SM}$ and compares them to lattice QCD calculations.  
The recent state-of-the-art lattice calculations of 
$R_{EM}$ and $R_{SM}$~\cite{Ale05} use a {\it linear}, 
in the quark mass ($m_q\propto m_\pi^2$), {\it extrapolation}
to the physical point,  
thus assuming that the non-analytic $m_q$-dependencies are  negligible. 
The thus obtained value for $R_{SM}$ at the physical 
$m_\pi$ value displays a large 
discrepancy with the  experimental result, as seen in Fig.~\ref{ratios}. 
The relativistic \ceft\ calculation, 
on the other hand, shows  that the non-analytic dependencies 
are {\it not} negligible. While
at larger values of $m_\pi$, 
where the $\Delta$ is stable, the ratios display a smooth 
$m_\pi$ dependence, at $m_\pi =\De $ there is an inflection point, and 
for  $m_\pi \leq \Delta$ the non-analytic effects are crucial.

One also notices from Fig.~\ref{ratios} that 
there is only little difference between the \ceft\ 
calculations with the $m_\pi$-dependence of 
$M_N$ and $M_\Delta$ accounted for, and an earlier calculation
\cite{Pascalutsa:2005ts}, where the ratios were evaluated 
neglecting the $m_\pi$-dependence of the masses. 

Fig.~\ref{ratios} also shows a theoretical uncertainty of the 
ratios $R_{EM}$ and $R_{SM}$ 
taken over the range  of $m_\pi^2 $ from 0 to 0.15~GeV$^2$. 
The $m_\pi$ dependence obtained from \ceft\  clearly shows that
the lattice results for $R_{SM}$ may in fact be consistent with experiment.

\subsection{$\Delta(1232)$ magnetic dipole moment}

Although the $\De(1232)$-isobar is the most distinguished and well-studied 
nucleon resonance, such a fundamental property as
its  magnetic dipole moment (MDM) has thusfar escaped a
precise  determination.
The problem is generic to any unstable particle whose lifetime
is too short for its MDM to be measurable in the usual way through 
spin precession experiments. A measurement of the MDM of
such an unstable particle can apparently be done only indirectly, in a 
three-step process, where the particle is first produced, then
emits a low-energy photon which plays the role of an external magnetic field, 
and finally decays. 
In this way the MDM of $\Delta^{++}$ is accessed in  
the reaction $\pi^+ p \to \pi^+ p \gamma$~\cite{Nef78,Bos91} while the
MDM of  $\Delta^+$ can be determined using the radiative pion photoproduction 
($\gamma p \to \pi^0 p \gamma^\prime$)~\cite{Drechsel:2000um}. 

A first experiment devoted to the MDM of $\De^+$ 
was completed in 2002~\cite{Kotulla:2002cg}. 
The value extracted in this experiment,
$\mu_{\Delta^+} =  2.7 {{+1.0} \atop {-1.3}}
(\mathrm{stat.}) \pm 1.5 (\mathrm{syst.}) \pm 3 (\mathrm{theor.}) $ [nuclear magnetons],   
 is based on theoretical input from 
the phenomenological model~\cite{Drechsel:2001qu,Chiang:2004pw}
 of the $\gamma p \to \pi^0 p \gamma^\prime$  reaction. 
To improve upon the precision of this measurement, 
a dedicated series of experiments has recently 
been carried out by the Crystal Ball Collaboration at MAMI~\cite{CB}. 
These experiments achieve about two orders of magnitude better 
statistics than the pioneering experiment\cite{Kotulla:2002cg}. 
The aim of the investigation within \ceft\ was to complement 
these high-precision measurements with an accurate 
and model-independent analysis of the 
$\gamma p \to \pi^0 p \gamma^\prime$ reaction.

\begin{figure}
\resizebox{0.45\textwidth}{!}{%
  \includegraphics{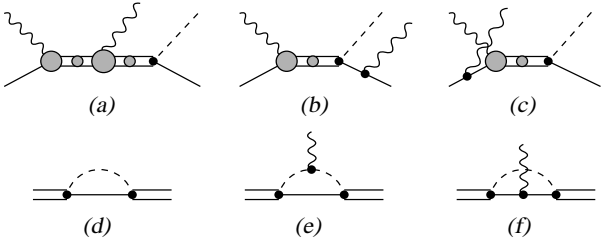} 
}
\caption{Diagrams for the $\gamma p \to \pi^0 p \gamma^\prime$ reaction 
at NLO in the $\delta$-expansion, considered in this work. Double lines represent
the $\De$ propagators.
}
\label{diagrams}
\end{figure}

\begin{figure}
\resizebox{0.4\textwidth}{!}{%
\includegraphics{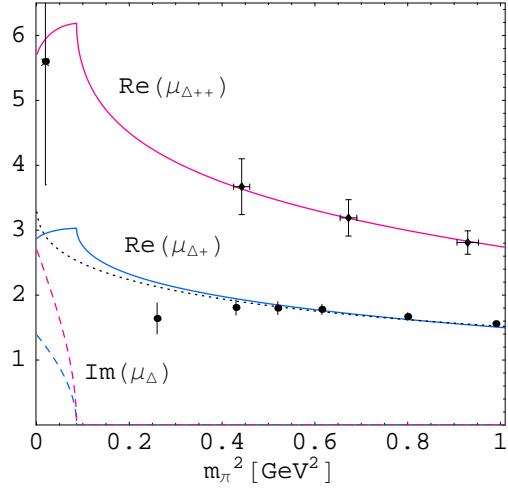}
}
\caption{Pion mass dependence of the real (solid curves) and imaginary
(dashed curves) parts of $\Delta^{++}$ and $\Delta^{+}$ MDMs [in nuclear magnetons].
Dotted curve is the result for the proton magnetic 
moment from Ref.~\cite{Pascalutsa:2004ga}. The experimental data point for $\De^{++}$
is from PDG analysis~\cite{PDG02}. Lattice data are from~\cite{Lein91} for $\De^{++}$ 
and  from~\cite{Lee} for $\De^+$.  
}
\label{chiral}
\end{figure}

The optimal sensitivity of the $\gamma p \to \pi^0 p \gamma^\prime$ reaction 
to the MDM term is achieved when the incident photon energy
is in the vicinity of $\De$, while the outgoing photon energy is 
of order of $m_\pi$. 
In this case the  $\gamma p \to \pi^0 p \gamma^\prime$ amplitude 
to next-to-leading order (NLO) in the $\de$-expansion 
is given by the diagrams of Fig.~\ref{diagrams}(a - c),
where the shaded blobs, in addition to vertices from 
Eqs.~(\ref{lagran1},\ref{lagran2},\ref{lagran3}),
 contain the one-loop corrections shown in Fig.~\ref{diagrams}(d - f).
The contributions to $\mu_\De$ of diagrams $(e)$ and $(f)$ 
in Fig.~\ref{diagrams} 
have been calculated in Ref.~\cite{PV05}, to which we refer for 
technical details. 
The evaluation of these loop diagrams also allows to quantify the 
$m_\pi$ dependence of $\mu_\De$ which can be used to compare 
with lattice QCD results. 
As all lattice data for $\mu_\Delta$ at present and in the foreseeable 
future are for larger than the physical values of $m_\pi$, 
their comparison with experiment
requires the knowledge of the $m_\pi$-dependence for this quantity.
Fig.~\ref{chiral} shows the pion mass dependence of real and
imaginary  parts of the $\Delta^+$ and $\Delta^{++}$ MDMs, according to 
our one-loop calculation. Each of the two solid curves has a free 
parameter, the counterterm $\mu_\De$ from $\lag^{(2)}_{N\De}$,
adjusted to agree with the lattice data at larger values of $m_\pi$.
As can be seen from 
Fig.~\ref{chiral}, the $\Delta$ MDM develops an 
imaginary part when $m_\pi<\Delta = M_\Delta - M$, 
whereas the real part has a pronounced cusp at $m_\pi = \Delta$. 
For $\mu_{\De^+}$, the curve 
is in disagreement with the trend of the recent lattice data,
which possibly is due to the ``quenching'' in the lattice calculations. 
The dotted line in Fig.~\ref{chiral} shows the result~\cite{Pascalutsa:2004ga}
for the magnetic moment for the proton. One sees that 
$\mu_{\Delta^+}$ and $\mu_p$, while having very distinct behavior
for small $m_\pi$, are approximately equal for larger values of $m_\pi$. 

\begin{figure}
\resizebox{0.4\textwidth}{!}{%
\includegraphics{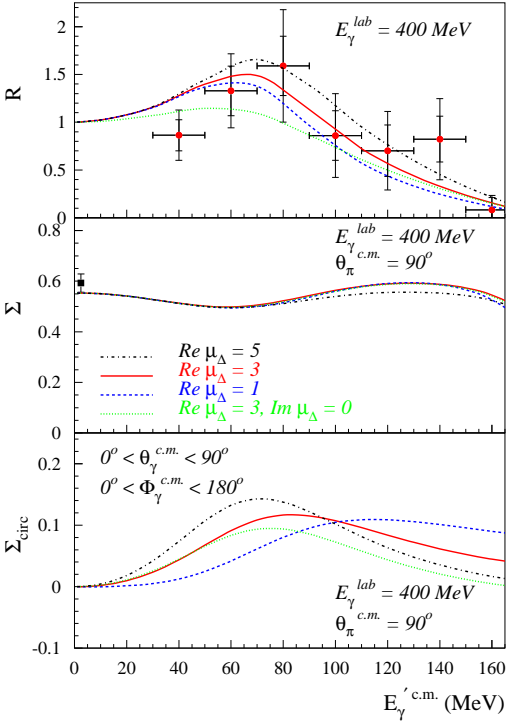}
}
\caption{The outgoing photon energy dependence of the 
$\gamma p \to \pi^0 p\ga'$ observables for different values of $\mu_{\Delta^+}$ 
(in units $e/2 M_\Delta$). 
Top panel: the ratio of $\gamma p \to \pi^0 p \gamma^\prime$ to 
$\gamma p \to \pi^0 p$ cross-sections Eq.~(\ref{eq:R1}). 
Data points are from~\cite{Kotulla:2002cg}. 
Middle panel: the linear-polarization photon asymmetry 
of the $\gamma p \to \pi^0 p \gamma^\prime$  
cross-sections differential 
w.r.t.\ the outgoing photon energy and pion c.m. angle. 
The data point at $E^\prime_\gamma = 0$ corresponds with the 
$\gamma p \to \pi^0 p$ photon asymmetry from~\cite{Beck:1999ge}.
Lower panel: the circular-polarization photon 
asymmetry (as defined in~\cite{Chiang:2004pw}), 
where the outgoing photon angles have been integrated over the 
indicated range. 
}
\label{fig:cross}
\end{figure}

We next discuss the \ceft\ 
results for the $\gamma p \to \pi^0 p \gamma^\prime$ 
observables. The NLO calculation of this process 
in the $\delta$-expansion corresponds with the diagrams of 
Fig.~\ref{diagrams}. 
This calculation completely fixes the imaginary  part of the 
$\gamma \Delta \Delta$ vertex. It leaves $\mu_\De$ as only free parameter,
which enters as a low energy constant in $\lag^{(2)}_{N \De}$. 
Thus the real part of $\mu_{\Delta^+}$ 
is to be extracted from the $\gamma p \to \pi^0 p \gamma^\prime$ observables, some of 
which are shown in Fig.~\ref{fig:cross} for an incoming photon 
energy $E_\gamma^{lab} = 400$~MeV as function of the emitted photon 
energy $E_\gamma^{\prime \, c.m.}$. In the soft-photon limit 
($E_\gamma^{\prime \, c.m.} \to 0$), the 
$\gamma p \to \pi^0 p \gamma^\prime$ 
reaction is completely determined from the bremsstrahlung process 
from the initial and final protons.  
The deviations of the $\gamma p \to \pi^0 p \gamma^\prime$ observables, 
away from the soft-photon limit, will then allow to study the 
sensitivity to $\mu_{\Delta^+}$. It is therefore very useful to 
introduce the ratio~\cite{Chiang:2004pw}:
\begin{eqnarray}
\label{eq:R1}
  R \,\equiv \, \frac{1}{\sigma_\pi} \cdot
  E^\prime_\gamma
  \frac{d\sigma}{dE^\prime_\gamma} ,
\end{eqnarray}
where $d\sigma / dE^\prime_\gamma$ is the 
$\gamma p \to \pi^0 p \gamma^\prime$ 
cross section integrated over the pion and photon angles, and 
$\sigma_\pi$ is the angular integrated cross section 
for the $\gamma p \to \pi^0 p$ process weighted with the bremsstrahlung 
factor, as detailed in~\cite{Chiang:2004pw}. 
This ratio $R$ has the property that in the soft-photon limit, the 
low energy theorem predicts that $R \to 1$. From 
Fig.~\ref{fig:cross} one then sees that the \ceft \ 
calculation obeys this theorem. This is a consequence of 
gauge-invariance which is maintained exactly throughout
the calculation, also away from the soft-photon limit.
 
The \ceft \ result for $R$ shows clear deviations 
from unity at higher outgoing photon energies, in good 
agreement with the first data for this process~\cite{Kotulla:2002cg}. 
The sensitivity of the \ceft \ calculation to the $\mu_\Delta$ is a very 
promising setting for the dedicated second-generation experiment which 
has recently been completed by the Crystal Ball Coll.\ at MAMI 
\cite{CB}. It improves upon the statistics of the first experiment 
(Fig.~\ref{fig:cross}) by at least two orders of magnitude 
and will allow for a reliable extraction of 
$\mu_{\Delta^+}$ using the \ceft \ calculation presented here. 
\newline
\indent
Besides the cross section, the $\gamma p \to \pi^0 p \gamma^\prime$ 
asymmetries for linearly and circularly polarized incident photons 
have also been measured in the recent dedicated experiment~\cite{CB}. 
They are also shown in Fig.~\ref{fig:cross}. 
The photon asymmetry for linearly polarized photons,  $\Sigma$, at $E^\prime_\gamma = 0$
exactly reduces  to the 
$\gamma p \to \pi^0 p$ asymmetry. It is seen from Fig.~\ref{fig:cross} that 
the \ceft \ calculation is in good agreement with the experimental value. 
At higher outgoing photon energies, the photon asymmetry
as predicted by the NLO \ceft \ calculation 
remains nearly constant and is very weakly dependent on $\mu_\Delta$. 
It is an ideal observable for a consistency check of the \ceft \ calculation 
and to test that the $\Delta$ diagrams of 
Fig.~\ref{diagrams} indeed dominate the process. 
Mechanisms involving $\pi$-photoproduction
Born terms followed by $\pi N$ rescattering have been considered in model 
calculations~\cite{Drechsel:2001qu,Chiang:2004pw}. In the $\de$-counting they
start contributing at next-next-to-leading order and therefore will provide 
the main source of corrections to the present NLO results.
\newline
\indent
The asymmetry for circularly polarized photons, 
$\Sigma_{circ}$, (which is exactly zero for a 
two body process due to reflection symmetry w.r.t.\ the reaction plane) 
has been proposed~\cite{Chiang:2004pw} as a unique observable 
to enhance the sensitivity to $\mu_\Delta$. 
Indeed, in the soft-photon limit , where the 
$\gamma p \to \pi^0 p \gamma^\prime$ process reduces to a two-body process, 
$\Sigma_{circ}$ is exactly zero. 
Therefore, its value at higher outgoing photon energies 
is directly proportional to $\mu_\Delta$. 
One sees from Fig.~\ref{fig:cross} (lower panel) 
that our \ceft \ calculation 
supports this observation, and shows sizeably different asymmetries 
for different values of $\mu_\Delta$. 
A combined fit of all three observables shown in Fig.~\ref{fig:cross} 
will therefore allow for a very stringent test 
of the \ceft \ calculation, which can then be used 
to extract the $\Delta^+$ MDM.

\section{Conclusions}
\label{sec:concl}

It was discussed here how Compton scattering sum rules relate 
low-energy nucleon structure quantities to the nucleon 
excitation spectrum, with special emphasis on the GDH sum rule. 
I demonstrated the utility of taking derivatives of the GDH
sum rule, in order to convert it to forms which are sometimes
more calculationally robust.  In particular it was shown how it allows  
to estimate the chiral extrapolations of lattice QCD results for 
anomalous magnetic moments of nucleons. 

Subsequently, new developments in our description of the nucleon 
excitation spectrum were discussed. In particular I reviewed recent 
work on a \ceft \ framework for the $\Delta(1232)$-resonance region.
This framework plays a dual role, in that it allows for
an extraction of resonance parameters from observables {\em and} 
predicts their  $m_\pi$ dependence. In this way it may provide
 a crucial connection of present lattice QCD results obtained at 
unphysical values of $m_\pi$ to the experiment.
This was demonstrated here explicitely for the $N$ and $\Delta$ 
masses, the $\gamma N \Delta$ transition and the $\Delta$ magnetic 
dipole moment. As the  next-generation lattice 
calculations of these quantities are on the way~\cite{Alexandrou:2005em}, 
such a \ceft\ framework will, hopefully, 
complement these efforts.

\section*{Acknowledgments}

I am grateful to my colleagues in Mainz for the unique 
culture of ``cross-fertilization'' between experiment and theory.
On the subject of two-photon physics, I like to thank in particular Dieter 
Drechsel and Barbara Pasquini, for the many collaborations. 
I also like to acknowledge Vladimir Pascalutsa for 
very fruitful recent collaborations 
on the \ceft \ in the $\Delta$-resonance region.  
This work is supported in part by DOE grant no.\
DE-FG02-04ER41302 and contract DE-AC05-84ER-40150 under
which SURA operates the Jefferson Laboratory.

\end{document}